\documentstyle[aps,prl,twocolumn,psfig]{revtex}
\draft

\begin{document}
\bibliographystyle{prsty}

\title{The Debye-Waller factor of liquid silica: Theory and simulation}
\author{Francesco Sciortino$^*$ and Walter Kob$^\dag$}

\address{$^*$Dipartimento di Fisica and Istituto Nazionale
per la Fisica della Materia, Universit\'a di Roma {\it La Sapienza},
P.le Aldo Moro 2, I-00185 Roma, Italy.}

\address{$^\dag$Institute of Physics, Johannes Gutenberg-University,
Staudinger Weg 7, D--55099 Mainz, Germany}
\date{24. June, 2000}
\maketitle

\begin{abstract}
We show that the prediction of mode-coupling theory for a model of a
network-forming strong glass-former correctly describes the wave-vector
dependence of the Debye-Waller factor. To obtain a good description it
is important to take into account the  triplet correlation function ${\bf
c}_3$, which we evaluate from a computer simulation.  Our results support
the possibility that this theory is able to accurately describe the
non-ergodicity parameters of simple as well as of network-forming liquids.

\end{abstract}

\vspace{-0.3cm}

The quantitative description of the glassy dynamics in liquids is
one important goal of modern research in condensed matter.  Work in
the last decade has provided evidence~\cite{vigo98}
that the so-called mode-coupling theory (MCT)~\cite{gotze99} is
able to describe the slow dynamics of {\it fragile} liquids in the weakly
supercooled state. Detailed theoretical predictions
for model systems --- including hard sphere systems, simple binary
liquids, and molecular liquids --- have been found to be in remarkable 
agreement with experimental measurements as well as with simulation 
results~\cite{barrat90,megen94,nauroth97,fabbian99,winkler00}. E.g. the
full $q$-dependence of the Lamb-M\"ossbauer and of the Debye-Waller
factors is predicted well by the theory. 

Recently, it has been shown that also intermediate
and strong glass-formers, such as glycerol or silica
(SiO$_2$)~\cite{franosch97,horbach99}, show features that are in
{\it qualitative} agreement with the predictions of MCT. However,
in the case of silica, a detailed comparison between theory
and numerical data has questioned the ability of MCT to describe
correctly the Debye-Waller factors of this important network-forming
material~\cite{horbach_diss,nauroth_diss}. In this Letter we show that
the disagreement between MD-results and theoretical predictions --- for
the case of silica --- were not due to failure of MCT to describe caging
in this network forming liquid but to a further approximation which is
assumed for the sake of simplicity in the commonly-used MCT equations.
We show that, once this approximation is avoided, MCT is able to describe
accurately the cages in silica, opening the way for a full description of
dynamics in network forming liquids above the critical temperature of MCT.

To start we briefly review the MCT equations.  The central quantity is
the coherent intermediate scattering function ${\bf F}({\bf q},t)$,
which for binary system can be written as a $2\times 2$ matrix
with entries $F_{ij}({\bf q},t)=\langle \delta\rho_i({\bf q},0)
\delta\rho_j^{\star}({\bf q},t)\rangle$. Here $\rho_i({\bf q},t)$ is
the density fluctuations for wave-vector ${\bf q}$, and $i$ is the label
for the species.

The equation of motion for ${\bf F}({\bf q},t)$ is given by
 
\begin{equation}
\!\!\!\ddot{{\bf F}}({\bf q},t)\!  +  \!
{\mbox{\boldmath $\Omega $}}^2(q){\bf F}({\bf q},t)+
\int_0^t \!\!d\tau {\bf M}({\bf q},t-\tau) \dot{{\bf F}}({\bf q},\tau) = 0,
\label{eq2}
\end{equation}
   
\noindent
where the frequency matrix is given by 
$\left[{\mbox{\boldmath $\Omega $}}^2(q)\right]_{ij}=q^2k_B T
(x_i/m_i)\sum_{k}\delta_{ik} \left[{\bf S}^{-1}(q)\right]_{kj}$. Here
$q=|{\bf q}|$, $x_i$ is the concentration of species $i$, $m_i$ is
their mass, and $\bf S$ is the partial structure factor matrix.  Within the
framework of MCT{\bf ,} the memory function $\bf M$ is given by $M_{ij}=x_i
k_B T N_{ij}/m_i$, where the matrix $N_{ij}$ is a quadratic form in
$F_{ij}({\bf q},t)$:

\begin{eqnarray}
N_{ij}({\bf q},t)& =& \frac{1}{2n x_i x_j}\int\frac{d
{\bf k}}{(2\pi)^3}
\sum_{\alpha\beta}\sum_{\alpha'\beta'}V_{i\alpha\beta}({\bf q},{\bf k}) \times\nonumber \\
& &V_{j\alpha'\beta'}({\bf q},{\bf q-k}) F_{\alpha\alpha'}({\bf k},t)
F_{\beta\beta'}({\bf q-k},t).
\label{eq3}
\end{eqnarray}

\noindent
Here $n$ is the particle density, and the vertices $V_{i\alpha\beta}
({\bf q},{\bf k})$ are given by

\begin{eqnarray}
V_{i\alpha\beta}({\bf q},{\bf k}) & = & \frac{{\bf q}\cdot {\bf
k}}{q}\delta_{i\beta} c_2^{i\alpha}({\bf k})+
\frac{{\bf q}\cdot ({\bf q}-{\bf k})}{q} \delta_{i\alpha} c_2^{i\beta}
({\bf q}-{\bf k})  \nonumber \\
& & \hspace{25mm}+n q c_3^{i\alpha\beta}({\bf q},{\bf q}-{\bf k}) \cdot
x_i .
\label{eq4}
\end{eqnarray}

\noindent
The function $c_2^{ij}({\bf q})=\delta_{ij}/x_i- \left[{\bf
S}^{-1}({\bf q})\right]_{ij}$ is the direct correlation function, and
$c_3^{i\alpha\beta}({\bf q},{\bf k})$ is the triple correlation function,
which is related to the triple density fluctuations via

\begin{eqnarray}
\langle \delta\rho_{\alpha}({\bf q}) \delta\rho_{\beta}({\bf k})
\delta\rho_{\gamma}^{\star}({\bf q}+{\bf k})\rangle & = &
N\sum_{\epsilon \sigma \eta}
S_{\alpha \epsilon} (q)
S_{\beta \sigma} (k)\nonumber \\
\times S_{\gamma \eta} (|{\bf q}+{\bf k}|) & &
\hspace{-5mm}
(\delta_{\epsilon \sigma}
\delta_{\epsilon \eta}
\delta_{\eta \sigma}/x_{\epsilon}^2 +
n^2 c_3^{\epsilon\sigma\eta}({\bf q},{\bf k})).
\label{eq6}
\end{eqnarray}

The only input required by MCT are the {\it static} quantities ${\bf S}$,
${\bf c}_3$, and $n$. Temperature enters the equations only through the
explicit $T$-dependence of ${\mbox{\boldmath $\Omega $}}^2(q)$  and
the implicit $T$-dependence of the static quantities.  The solution
of this type of equation shows at low $T$ a two-step relaxation
dynamics, if $T$ is above $T_c$, the so-called critical temperature of
MCT~\cite{gotze99}. Below $T_c$ the time correlation functions do not
decay to zero anymore, thus signaling that the system is not longer
ergodic. Hence the height of the plateau in a time correlation function
is usually called the non-ergodicity parameter (NEP) since it measures
that fraction of the correlation that does not decay to zero even at
long times. The physical relevance of the NEP can be inferred from the
fact that it can be directly measured in light and neutron scattering
experiments. For the case of $F({\bf q},t)$ the NEP is ${\bf F}_c(q)$,
the Debye-Waller factor for wave-vector $q$.

In one of the first attempts to solve the MCT equations for soft
spheres, the possibility of setting in Eq.~(\ref{eq4}) ${\bf c}_3=0$
was considered~\cite{barrat90}.  This approximation corresponds to a
factorization of the triple density correlation in $q$-space in product
of the three pair density correlation.  It was found that for this
simple liquid, this approximation does not significantly affect the
MCT predictions.  All further works have build on this information.
In this Letter we check the validity of this approximation for
network-forming liquids.  We use molecular dynamics (MD) simulations to
determine the triple correlation function for silica\cite{bkspotential}
and use these functions to solve the full MCT equations. This calculation
allows us to determine for the first time whether or not MCT is able to
give a quantitative description of the cages of {\it strong} glass-formers.
For completeness, we perform the same study for a well studied binary
mixture of Lennard-Jones particles (BMLJ)\cite{bmljpotential}.

In the past, the dynamics of both systems has been
carefully analyzed and it has been shown that the relaxation
dynamics shows the qualitative features predicted by the
MCT~\cite{horbach99,horbach_diss,nauroth_diss,kob95,gleim98}. For
the case of the BMLJ also a quantitative comparison has been made
in that ${\bf F}_c(q)$ has been calculated theoretically (assuming
${\bf c}_3=0$) and compared with simulation data.  This comparison
showed that very good agreement between theory and simulation is
obtained, even if ${\bf c}_3=0$ is assumed, in agreement with the
conclusions of Barrat and Latz~\cite{barrat90}. But for the case
of SiO$_2$ a similar comparison showed that the MCT prediction for
${\bf F}_c(q)$ does not describe properly the confining cage if ${\bf
c}_3=0$~\cite{horbach_diss,nauroth_diss}.

In order to investigate the influence of ${\bf c}_3$ it is necessary
to determine $c_3^{\epsilon\sigma\eta}({\bf q},{\bf k})$ with high
precision.  For this we calculated $\langle \delta\rho_{\alpha}({\bf k})
\delta\rho_{\beta}({\bf p}) \delta\rho_{\gamma}^{\star}({\bf q}) \rangle$
with ${\bf q}={\bf k}+{\bf p}$ for all triplets of values $|\bf k|$,$|\bf
p|$,$|\bf q|$, with $|\bf p| \ge |\bf k|$ and $|{\bf p}-{\bf k}| < |\bf
q| < | {\bf p}+{\bf k}|$, and then used Eq.~(\ref{eq6}) to determine
${\bf c}_3$. We have selected a mesh size $\Delta$, where $\Delta$
was $0.154$\AA$^{-1}$ and $0.334$ for the case of SiO$_2$ and BMLJ,
respectively. Only wavevectors with modulus less than $100\Delta$ have
been considered.  For each triplet of value for $|\bf k|$,$|\bf p|$,
$|\bf q|$ we randomly selected up to $10^2$ combinations of wave-vectors
$\bf k$, $\bf p$, $\bf q$, calculated for these the density fluctuations
and then the triple density correlation function. For BMLJ this was
done for 12309 completely independent configurations, corresponding to
100 million time steps in the simulation. For SiO$_2$ we analyzed
8577 configurations (20 million time steps).  Note that this very large
number of independent configurations was needed in order to determine
${\bf c}_3$ with sufficient high accuracy to make sure that the final
results for the NEP do not depend anymore on the noise in ${\bf c}_3$.
We mention that using only 1000 independent configurations was, e.g.,
not sufficient to guarantee this.  Due to the large effort needed to
generate the configurations and to analyze them, we determined ${\bf c}_3$
only for one temperature, namely $T=1.0$ (BMLJ) and $T=4000$K SiO$_2$.
Thus in the calculation of the MCT vertices in Eq.~(\ref{eq4}) we assumed
that the direct correlation functions depend on temperature but that
the $T$-dependence of ${\bf c}_3$ can be neglected.

\begin{figure}[b]   
\psfig{file=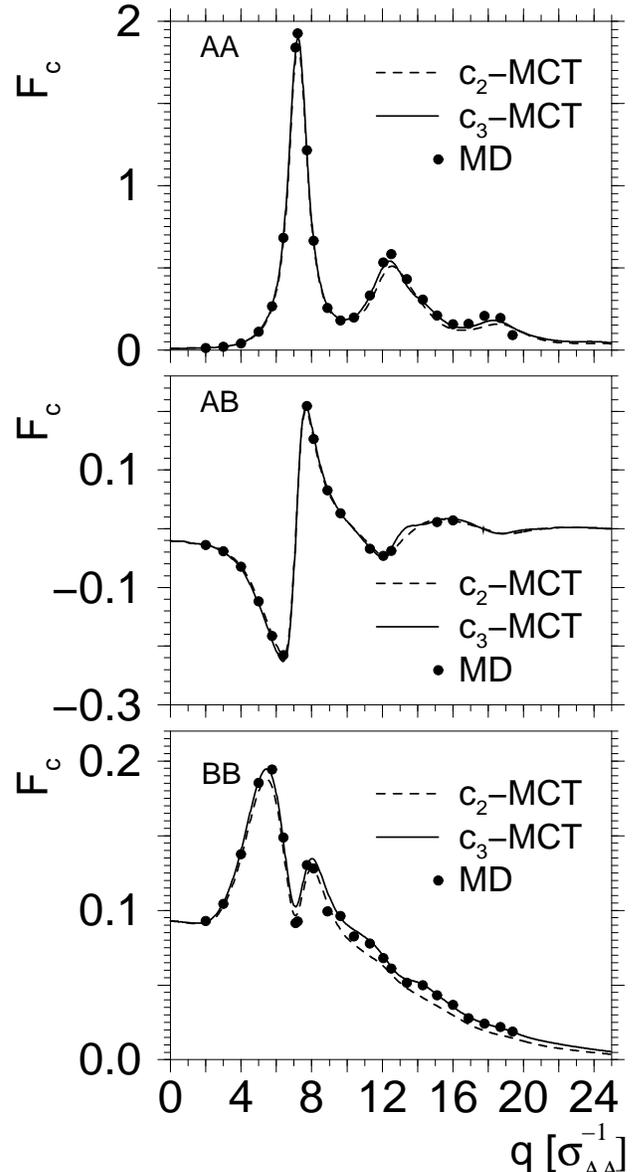,width=9cm,height=15.5cm}
\vspace{2mm}
\caption{
Wave-vector dependence for the non-ergodicity parameters for
the BMLJ system. The solid and dashed curves are the theoretical
prediction with and without the inclusion of the ${\bf c}_3$
terms in Eq.~\protect(\ref{eq3}).  The circles are the MD results
from~\protect\cite{gleim98}.
}
\label{fig1}
\end{figure}

\begin{figure}
\psfig{file=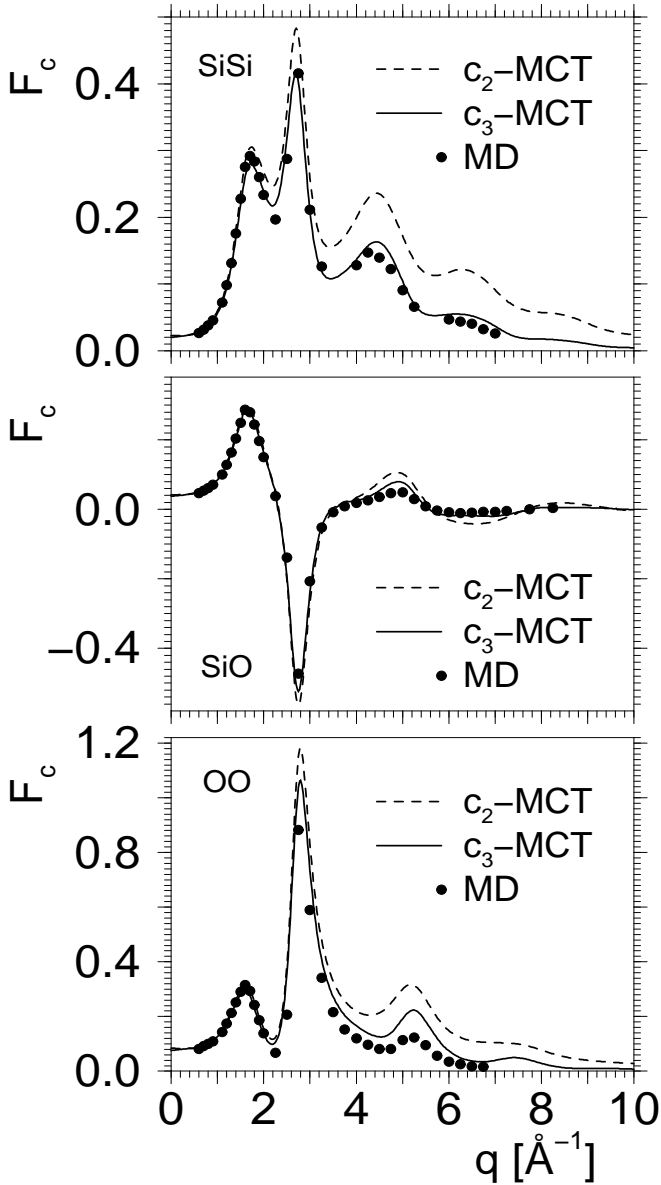,width=9cm,height=15.5cm}
\vspace{2mm}
\caption{The same quantities as in Fig.~\protect\ref{fig1}, but for
silica. The MD data is from Ref.~\protect\cite{horbach_diss}.}
\label{fig2}
\end{figure}

~From the equations of motion, Eqs.~(\ref{eq2})-(\ref{eq4}), we
determined $T_c$ and ${\bf F}_c(q)$, using the iteration procedure
from Ref.~\cite{barrat90}. The results obtained are presented
in Figs.~\ref{fig1} and \ref{fig2} were we show the three components
of the NEP for the case of the BMLJ and silica, respectively. In
each panel three curves are shown: the data from the simulation
(filled circles)~\cite{gleim98}, the theoretical prediction if
${\bf c}_3=0$ (curves labeled with ``$c_2$''), and the theoretical
prediction if ${\bf c}_3$  is taken into account (curves labeled with
``$c_3$'')~\cite{footnote}. From Fig.~\ref{fig1} we see that the
curves for $c_2$ and $c_3$ are very close together, from which we
conclude that in the case of the BMLJ the inclusion of ${\bf c}_3$
changes the prediction of the theory only weakly. The main difference
between the two curves is in the $BB$ correlation, since the $c_3$-curve
describes the simulation data even better than the $c_2$-curve. Also the
theoretical prediction for $T_c$ is essentially independent of whether or
not ${\bf c}_3$ is neglected, in that $T_c$ decreases from $T_c=0.922$
to $T_c=0.910$ if ${\bf c}_3$ is taken into account. These results are
in agreement with the findings of Ref.~\cite{barrat90}.

~From Fig.~\ref{fig2} we recognize that for the case of SiO$_2$ the
influence of ${\bf c}_3$ on the NEP is much stronger than for the BMLJ
system. For this network-forming liquid the inclusion of ${\bf c}_3$ into
the vertices leads to predicted ${\bf F}_c(q)$ which are in substantially
better agreement with the ones from the simulation than the ones from
the $c_2$-theory. The improvement is particularly noticeable for $q$
above the location of the first peak, a length scale that corresponds
to the distance between two neighboring tetrahedra. E.g. in the case of
the Si-Si correlation the relative difference between the theoretical
curve and the MD data is decreased by a factor around five. The remaining
difference is now only on the order of the error of the MD data. Note that
${\bf F}_c(q)$ provides information on the cage in which a particles is
confined and which is formed by its neighboring particles. The fact that
in the vicinity of the first maximum the two theoretical curves as well as
the MD data are very close together shows that already the $c_2$-theory
is able to capture the structure of the cage on this length scale. For
larger wave-vectors the $c_2$-curve is too high, which means that the
size of the cage is underestimated. Only if the terms due to ${\bf c}_3$
are taken into account, a reliable description for the cage is obtained.

For the Si-Si correlation the mentioned decrease of the NEP is more
pronounced than for the two other correlation functions, which is
reasonable since it is the Si-atoms that sit in the center of the
tetrahedra, i.e. that make up the network structure, and hence it can
be expected that the inclusion of the ${\bf c}_3$-terms is important.
Nevertheless, from the figure we see that the inclusion of these terms
lead also to a significant improvement of the theoretical prediction
for the NEPs for Si-O and O-O if $q$ is larger than the location of
the first peak. E.g. for the case of O-O the relative error of the
$c_3$-curve for $q\geq 2.0$\AA\mbox{ }is only half as large as the one
for the $c_2$-curve. Thus we conclude that the $c_2$-MCT is able to give
a good description of the cage for the oxygens only on the length scale
corresponding to the first maximum {\bf of $F(q,0)$}.

The strong influence of the ${\bf c}_3$-terms on ${\bf F}_c(q)$ is
also found in the value of $T_c$. If one sets ${\bf c}_3=0$ one finds
$T_c=3962$K, whereas the $c_3$-theory predicts $T_c=4676$K. Thus,
whereas for the BMLJ system we found only a change of the order
of 1\%, we now find a change around 18\%. We mention, however,
that MCT usually overestimates the value of $T_c$, and indeed we
have $T_c^{MD}=0.435$ and $T_c^{MD}=3330$K for BMLJ and SiO$_2$,
respectively~\cite{kob95,horbach99}.

\begin{figure}
\psfig{file=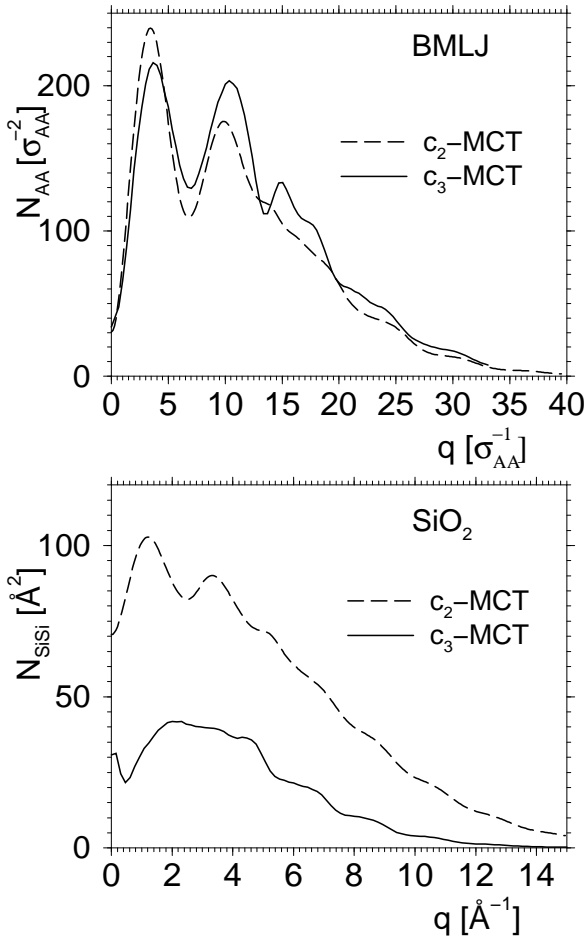,width=8cm,height=12.5cm}
\caption{Wave-vector dependence of the A-A and Si-Si component of the
memory function (upper and lower panel, respectively). The dashed and
solid curves are for the version of MCT in which the ${\bf c}_3$-terms are
neglected and taken into account, respectively.
}
\label{fig3}
\end{figure} 

Finally we show in Fig.~\ref{fig3} the $q$-dependence of $N_{\rm AA}$ and
$N_{\rm SiSi}$ from Eq.~(\ref{eq3}) for the corresponding $T_c$. From this
figure we see that for the BMLJ case the $c_2$-memory function is very
similar to the one in which ${\bf c}_3$ is taken into account and thus it
is not surprising that also the NEP and the $T_c$ are not changing. This
is in stark contrast to the SiO$_2$ system, for which the memory functions
from the $c_2$- and $c_3$-theory are very different. Note that in a binary
mixture each component of the memory function has contributions from 16
different terms (see Eq.~(\ref{eq3})).  We found that in the case of
SiO$_2$ the inclusion of the ${\bf c}_3$-terms leads to a significant
change of the contribution involving the off-diagonal terms. In the
$c_2$-case these contributions are {\it negative}, thus they drive the
transition to lower temperatures. If the ${\bf c}_3$-terms are taken
into account, these contributions become less negative and hence the
transition takes place already at higher temperatures.

In summary, the present calculation shows that MCT is able to give an
accurate {\it quantitative} description of the NEP for a strong
glass-former.  This opens the way for a detailed description of the
dynamics in the MCT region for this class of materials. Hence our results
support the possibility that MCT is able to make quantitative predictions
for all types of glass-formers.

Acknowledgment: We thanks L. Fabbian and M. Nauroth for contributing to
the early development of this research, and J. Horbach for providing
the SiO$_2$ NEPs. Part of this work was supported by the DFG through
SFB~262. F.S. acknowledges support from INFM-PRA-HOP and MURST-PRIN98.
W.~K. thanks the Universit\'a La Sapienza for a visiting professorship
during which part of this work was carried out.


\begin{references}
\bibitem{bmljpotential}
The BMLJ system is composed of 1000 particles, 80\% A and 20\% B
particles, with the same mass $m$, which interact via a Lennard-Jones
potential of the form $V_{\alpha\beta}(r)= 4\epsilon_{\alpha\beta}
[(\sigma_{\alpha\beta}/r)^{12}- (\sigma_{\alpha\beta} /r)^{6}]$, with
$\alpha, \beta \in\{{\rm A,B}\}$ and cut-off radii $r_{\alpha\beta}^{\rm
c}$=$2.5 \sigma_{\alpha\beta}$. The parameters are $\epsilon_{\rm
AA}=1.0$, $\sigma_{\rm AA}=1.0$, $\epsilon_{\rm AB}=1.5$, $\sigma_{\rm
AB}=0.8$, $\epsilon_{\rm BB}=0.5$, and $\sigma_{\rm BB}=0.88$. We use
$n=1.2$, with $\sigma_{\rm AA}$ as the unit of length.
\bibitem{footnote}
Note that ${\bf F}_c(q)$ are {\it not} normalized by the static 
structure factor, since this would give rise to singularities if  
$S_{\alpha\beta}(q)=0$.  To account for the $T$-dependence of ${\bf
F}(q,0)$, we first divide ${\bf F}_c(q)$ by ${\bf F}(q,0)$ at the    
corresponding $T_c$ and then multiply this function by ${\bf F}(q,0)$
of the $c_2$-curve.
\end{references}
\end{document}